\documentclass[12pt]{iopart}
 \usepackage{graphicx}
 \def\be{\begin{equation}}
 \def\ee{\end{equation}}
 \def\bea{\begin{eqnarray}}
 \def\eea{\end{eqnarray}}
 \def\p{\partial}
 \def\Im{{\rm Im}}
 \def\Re{{\rm Re}}
 \def\Nf{N_{\!f}}
 \def\gsim{\mathrel{\rlap{\lower0.2em\hbox{$\sim$}}\raise0.2em\hbox{$>$}}}
 \def\bm#1{\mbox{\boldmath$#1$}}
 
 \def\rr{\rule[-0.3em]{0em}{1.5em}}
\begin{document}

\title{Hard parton damping in hot QCD}

\author{Andr{\'e} Peshier}

\address{Institut f{\"u}r Theoretische Physik, Universit{\"a}t Giessen,
         35392 Giessen, Germany}

\ead{Andre.Peshier@theo.physik.uni-giessen.de}

\begin{abstract}
The gluon and quark collisional widths in hot QCD plasmas are discussed with emphasis on temperatures near $T_c$, where the coupling is large.
Considering the effect on the entropy, which is known from lattice calculations, it is argued that the width of the partons, which in the perturbative limit is given by $\gamma \sim g^2 \ln(g^{-1})\, T$, should be sizeable at intermediate temperatures but has to be small close to $T_c$.
This behavior implies a substantial reduction of the radiative energy loss of jets near $T_c$.
\end{abstract}

\section{Introduction}
The damping rate, or the collisional width, as well as the dispersion relation of excitations are important features of many-particle systems.
In hot QCD, the calculation of the width is complicated due to screening effects; already at leading order in the coupling the width is sensitive to the soft magnetic sector of QCD, which is not yet fully understood.
Even less is known in the nonperturbative regime: Schwinger-Dyson approaches have to face the issues of nonperturbative renormalization and gauge invariance, and addressing real-time properties within lattice QCD is nontrivial.

In an alternative approach it was proposed \cite{Peshi04} to infer essential 1-particle properties from suitable `known' quantities, i.\,e., taking a top-down view on the fact that certain quantities can be (rather directly) expressed in terms of the {\em full} propagator.
Here, the relation of the propagator to the entropy is outlined first for a scalar theory, before extending the results \cite{Peshi04} from quenched to full QCD. The findings should have interesting implications for various quantities as demonstrated here for the energy loss of hard jets.

\section{Spectral function and entropy}
The thermodynamic potential $\Omega$ of a system of particles with a given interaction can be expressed in terms of the exact 2-point function(s) \cite{LuttiW}. For a scalar theory with the (retarded) propagator $\Delta$ it can be written as (see, e.\,g., \cite{Peshi04})
\be
  \Omega
  =
  \int_{k^4} n_b(\omega)\,
  \Im\!\left( \ln(-\Delta^{\!-1}) + \Pi\Delta \right)
  -\Phi[\Delta]\, ,
\ee
where $\Pi = \Delta_0^{\!-1}-\Delta^{\!-1}$ is the self-energy, $\Phi$ is the sum of the 2-particle irreducible skeleton diagrams, $n_b(\omega) = (\exp(\omega/T)-1)^{-1}$ is the boson distribution function, and $\int_{k^4} = \int d\omega/(2\pi)\int_{k^3}$, $\int_{k^3} = \int d^3k/(2\pi)^3$.
The functional $\Omega[\Delta]$ is to be evaluated with the exact propagator which is obtained from the stationarity condition $\delta\Omega[\Delta]/\delta\Delta = 0$, or 
$\Pi = 2\,\delta\Phi / \delta\Delta$,
i.\,e., the self-energy follows diagrammatically by cutting a propagator line in the skeleton graphs of $\Phi$.
Taking $\delta\Omega/\delta\Delta = 0$ into account leads to the entropy,
\bea
  s
  &=&
  -\frac{\p\Omega}{\p T}
  \,=\,
  -\int_{k^4} \frac{\p n_b}{\p T}\, \Im\!\left(
    \ln(-\Delta^{\!-1}) + \Pi\Delta
   \right)
  +\left.\frac{\p\Phi}{\p T}\right|_{\Delta}
  \,=\,
  s^{dqp}+s' \, ,
  \nonumber
  \\
  &&
  s^{dqp}
  \,=\,
  -\int_{k^4} \frac{\p n_b}{\p T}
   \left(
    \Im\ln(-\Delta^{-1}) + \Im\Pi\, \Re\Delta 
   \right) ,
  \nonumber
  \\
  && 
  s'
  \,=\,
  -\int_{k^4} \frac{\p n_b}{\p T}\, \Re\Pi\, \Im\Delta
  +\left.\frac{\p\Phi}{\p T}\right|_{\Delta} .
\eea
From the exact expressions of $\Omega$ and $s$, self-consistent (`$\Phi$-derivable') approximations \cite{Baym} follow by truncating the expansions of $\Phi$ and, accordingly, of $\Pi$ at a given loop order. In terms of the `naive' perturbation theory based on the free propagator $\Delta_0$, this corresponds to a resummation of whole classes of diagrams. 
At 3-loop order, with
$\Phi
  =
  3\, \raisebox{-.5em}{\includegraphics[scale=0.6]{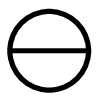}}
  +
  3\, \raisebox{-.5em}{\includegraphics[scale=0.6]{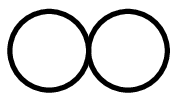}}
  +
  12\! \raisebox{-.6em}{\includegraphics[scale=0.4]{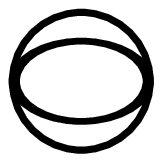}}\!$,
the term $s'$ vanishes, and the total entropy is given entirely by the so-called dynamical quasiparticle contribution $s^{dqp}$. For topological reasons this holds also in other theories, cf.~\cite{BlaizIR} and references therein, for $\Phi$-graphs with one and two vertices. 
This is interesting in particular with regard to QCD since it was argued \cite{Peshi02} that {\em leading}-loop resummations of thermodynamic quantities yield expedient approximations at large coupling. 
While seemingly counter-intuitive, this is due to the generally presumed asymptotic nature of perturbative expansions, which implies that higher order terms in the coupling $\alpha$ almost compensate each other numerically at an order which {\em decreases} with $\alpha$. On the other hand, {\em some} parts of higher order contributions in $\alpha$ have to be resumed for the thermodynamical consistency of the approximation, because the coupling satisfies a renormalization group equation.

It is noteworthy that the genuinely nonperturbative approximation $s \simeq s^{dqp}$ has a simple 1-loop structure and does not depend on the vertices. 
It can be decomposed,
\bea
  s
  &=&
  s^{(0)}+\Delta s \, ,
  \nonumber \\
  &&
  s^{(0)}
  =
  \frac1T
  \int_{k^3}
  \left( 
    -T\ln\left(1-e^{-\omega_k/T}\right) + \omega_k\, n_b(\omega_k) 
  \right) ,
  \nonumber
  \\
  &&
  \Delta s
  =
  \int_{k^4} \frac{dn_b}{dT}
  \left( {\rm arctan}\, \lambda - \frac\lambda{1+\lambda^2} \right) ,
  \label{eq: sdqp}
\eea
with $\lambda = \Im\Delta/\Re\Delta$ (and where from now on the superscript in $s^{dqp}$ is omitted for notational convenience). The contribution $s^{(0)}$ has the form of the entropy of an ideal gas with a dispersion relation $\omega_k$. However, $\omega_k$ is here determined from $\Re\Delta(\omega_k) = 0$ rather than being the real part of the pole (if existent) of the propagator, but for simplicity it is still referred to as the `dispersion relation'. 
The contribution $\Delta s$ is due to a nontrivial (imaginary part of the) propagator.
Since it is difficult to calculate even in a (self-consistent) approximation, the entropy functional (\ref{eq: sdqp}) is in the following evaluated with a physically motivated {\em Ansatz} for the propagator. 
To this end, the propagator is expressed in the Lehmann representation in terms of the spectral function,
\be
  \Delta(k_0,\bm k)
  =
  \int_{-\infty}^\infty \frac{d\omega}{2\pi}\,
    \frac{\rho(\omega,\bm k)}{k_0-\omega} \, .
\ee
It can then be shown on rather general grounds \cite{Peshi04} that the contribution $\Delta s$ to the entropy is positive. This is plausible since the entropy measures the population of the phase space, and a nontrivial $\rho = -2\Im\Delta$ describes particles off the mass shell.

An often used {\em Ansatz} to model non-zero width is obtained by replacing the free spectral function
$\rho_0(\omega)
  =
  2\pi \left[\delta((\omega - k)^2) - \delta( (\omega + k)^2 )\right]$
by a Lorentzian form,
\be
 \rho_L(\omega)
 =
 \frac\gamma{E} \left(
   \frac1{(\omega-E)^2+\gamma^2} - \frac1{(\omega+E)^2+\gamma^2}
 \right) ,
 \label{eq: rhoL}
\ee
where the parameter $E$ is related to the position of the peak of $\rho_L$, and $\gamma$ to its width. $E^2(\bm{k}) = \bm k^2+m^2-\gamma^2$ corresponds to parameterizing the real part of the self-energy by a (constant, see below) mass term $m^2$, which leads to the retarded propagator
\be
  \Delta_L(\omega, \bm k)
  =
  \frac1{\omega^2-\bm k^2-m^2+2i\gamma\omega} \, ,
  \label{eq: Delta_L}
\ee
and to a dispersion relation $\omega_m = (m^2+\bm k^2)^{1/2}$.
In the resulting entropy $s_L = s_L^{(0)}+\Delta s_L$,
\bea
  s_L^{(0)}(m)
  &=&
  \frac1T
  \int_{k^3} \left( 
    -T\ln(1-e^{-\omega_m/T}) + \omega_m\, n_b(\omega_m/T) 
  \right) ,
  \nonumber
  \\
  \Delta s_L(m,\gamma)
  &=&
  \int_{k^4}\frac{\partial n_b}{\partial T}\,
   \left(
   \arctan\frac{2\gamma\omega}{\omega_m^2-\omega^2}
   -\frac{2\gamma\omega(\omega_m^2-\omega^2)}
          {(\omega^2-\omega_m^2)^2+(2\gamma\omega)^2}
   \right) ,
   \label{eq: sL}
\eea
the first contribution is simply the entropy of free bosons with the mass $m$.
As expected, the contribution $\Delta s_L$ is positive. It turns out that the decreasing effect of the mass in $s_L(m,\gamma)$ can be compensated by the width; for $\gamma > m$, the entropy even exceeds the Stefan-Boltzmann value $s_0 = \frac4{90}\, \pi^2 T^3$, cf.~figure \ref{fig: sL(m,Gamma)}.
\begin{figure}[hb]
 \hskip -1em
 \includegraphics[scale=0.75]{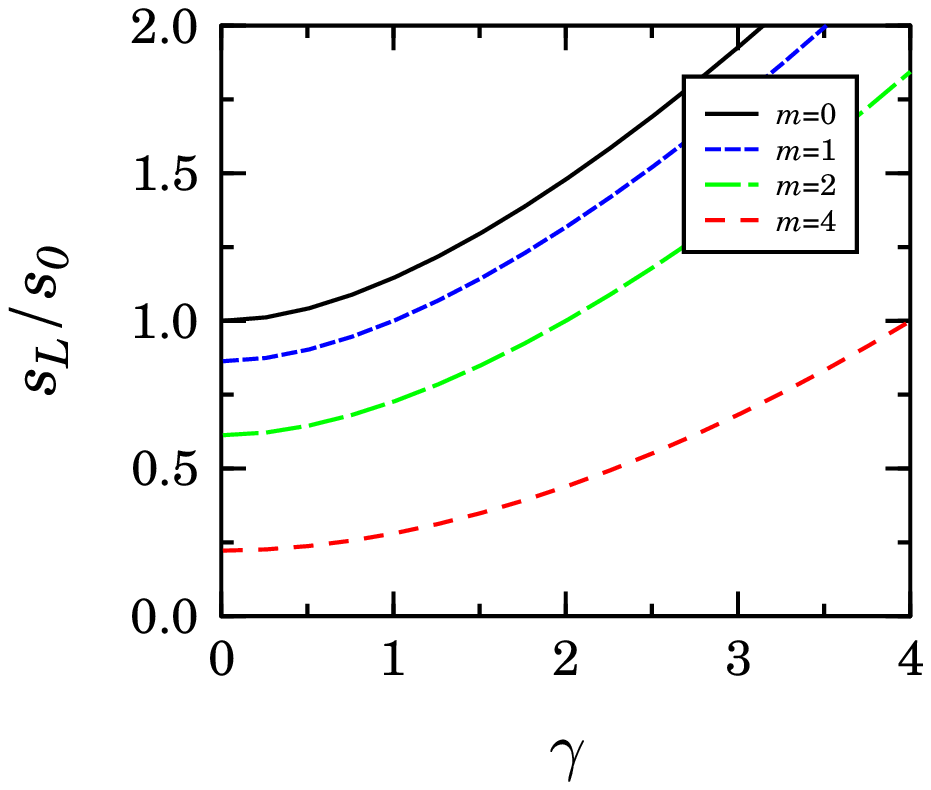}
 \hfill
 \raisebox{8mm}{
 \includegraphics[scale=0.55]{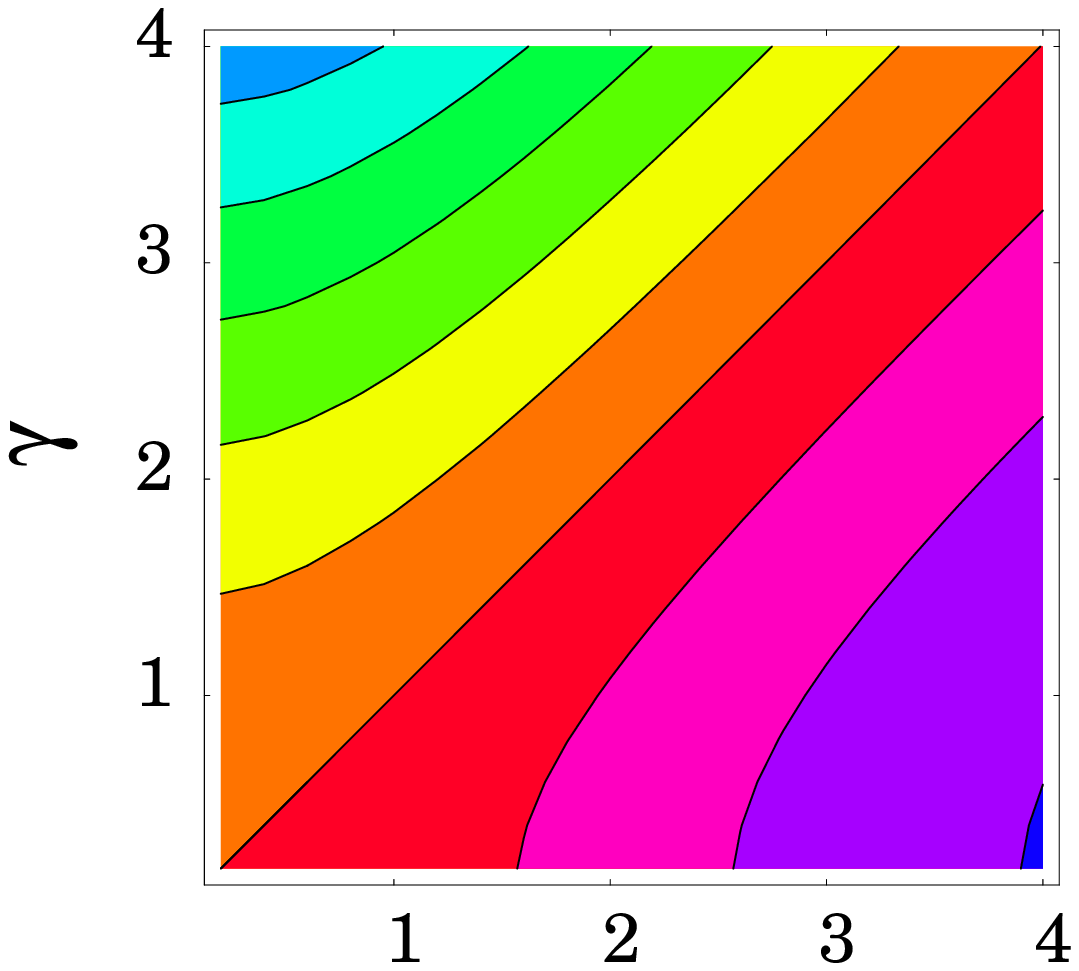}\rule{2em}{0em}}
 \caption{
   Left: The entropy $s_L(m,\gamma)$ as a function of $\gamma$ for several 
   values of $m$ (both in units of the temperature).
   Right: Contour plot of $s_L/s_0$; the contour spacing is 0.25, and $s_L$
   coincides with $s_0$ along the line $\gamma=m$, as proven in 
   \cite{Peshi04}.
  \label{fig: sL(m,Gamma)}}
\end{figure}

A few remarks are in order here. It should first be emphasized that the approach is not restricted to a small width, i.\,e., the excitations need not to be quasiparticles in the strict sense, with $\gamma \ll \omega_k$. 
Second, the quantities $m$ and $\gamma$ introduced above should be considered as parameters of the spectral function in the thermodynamically relevant part of the phase space, for hard momenta $k \sim T$. As a bulk property, the entropy is indeed rather insensitive to their behavior at other momenta.
The scope of the second assumption, the Lorentzian form of $\rho$, was studied by considering also other conceivable spectral functions. Without going to details, for which I refer again to \cite{Peshi04}, I mention that the entropy depends mainly on the long-time behavior of the Fourier transform $\rho(t)$, which for the Lorentzian case is $\rho_L(t) \sim \exp(-\gamma t)\sin(Et)$. It turns out that other classes of spectral functions show a comparable increase in the entropy with the inverse of the characteristic attenuation time of $\rho(t)$. An exception are hypothetical models with $\rho(t) \sim (\gamma t)^{-1}$, where the enhancement is minimal. Such a power-law behavior, however, is not expected for physically relevant theories.\footnote[8]{In hot QED, e.\,g., the fermion propagator is an {\em 
    entire} function of the energy \cite{BlaizI}. 
    Nonetheless, the spectral function is strongly peaked, with a 
    characteristic width $\sim e^2 \ln(e^{-1})T$. Since in Fourier space it
    decreases faster than an exponential, the effect for the entropy will 
    be even more pronounced than for a Lorentzian spectral function.}
Therefore, a seizable increase of the entropy with the width of the excitations should be a generic effect in interacting many-particle systems. In QCD, this will put a restriction on the width of hard partons.

\section{Parton width and entropy in QCD}

The approach outlined above for a scalar theory can be easily applied to hot QCD. The gluon propagator has a transverse and a longitudinal part; similarly, the quark propagator describes a particle and a hole excitation. The longitudinal and the hole excitations are collective modes which give only minor contributions to the entropy, which are neglected here.\footnote[3]{
    More precisely, they are of order $g^3$ in perturbation theory. For 
    larger coupling, in an approximately self-consistent approach based on 
    the hard-thermal-loop propagators \cite{BlaizIR, Peshi01}, they are 
    numerically small.}
Accounting for the degeneracies, for $\Nf$ flavors, the entropy
\bea
  s^{QCD}
  &=&
  -2(N_c^2-1)
  \int_{k^4} \frac{\partial n_b}{\partial T}\, \left( 
    \Im\ln(-\Delta^{-1}) + \Im\Pi\, \Re\Delta \rule{0em}{1.2em} 
  \right) 
  \nonumber \\
  && 
  -2N_c \Nf
  \int_{k^4} \frac{\partial n_f}{\partial T}\,
   \left( \Im\ln(-S^{-1}) + \Im\Sigma\, \Re S \rule{0em}{1.2em} \right)
 \label{eq: s_QCD}
\eea
is thus expressed in terms of the transverse gluon propagator $\Delta$ and the quark particle propagator $S$;\footnote[7]{$\Delta$ and $S$ denote here 
    the scalar coefficients of the respective contributions to the gluon or
    quark propagator, cf.~\cite{Peshi01} for conventions. In particular, 
    the particle contribution to the quark propagator follows by multiplying
    $S(K)=(K^2-\Sigma)^{-1}$ with a spinor constructed from the 4-momentum
    $K$.}
here $n_f(\omega) = (\exp(\omega/T)+1)^{-1}$ is the fermion distribution function.

In a first model for the propagators only the real contributions to the self-energies may be considered, which amounts to gauge invariant mass terms in the gluon and quark propagators,
\be
  m_g^2
  =
  \frac16 \left( N_c+ \textstyle\frac12\, \Nf \right) T^2 g^2 \; ,
  \qquad
  m_q^2
  =
  \frac{N_c^2-1}{8N_c}\, T^2 g^2 \, .
\ee
The resulting entropy, containing then only the contributions $s^{(0)}$, cf.~(\ref{eq: sL}), is a thermodynamical consistent resummation, and with the coupling parameterized as
\be
  g^2(T)
  =
  \frac{48\pi^2}{(11N_c-2\Nf)\ln\left( \frac{T-T_s}{T_c/\lambda} \right)^2} \, ,
  \label{eq: g2}
\ee
the `effective' quasiparticle (eQP) model \cite{eQP} can describe lattice data for various $\Nf$.

It is more physical, however, to take into account the width of the excitations. The results derived in \cite{Pisar93} lead to a parameterization of the gluon and the quark width,
\be
  \gamma_g
  =
  N_c\, \frac{g^2T}{8\pi}\, \ln\frac{c}{g^2}\; ,
  \qquad
  \gamma_q
  =
  \frac{N_c^2-1}{2N_c}\, \frac{g^2T}{8\pi}\, \ln\frac{c}{g^2} \, .
 \label{eq: gamma}
\ee
It is emphasized that the functional form of the masses and the widths is given, hence a description of the lattice data by the entropy (\ref{eq: s_QCD}), with Lorentzian propagators, is nontrivial.
For quenched QCD, it was shown in \cite{Peshi04} that this dynamical quasiparticle (dQP) approach actually improves the fits compared to the effective quasiparticle model \cite{eQP} without width.
For temperatures $T \gsim 1.2T_c$, the width was found to be of a similar size as the mass, hence the excitations are, in fact, not quasiparticles in the strict meaning. This could be expected from the general parametric behavior $\gamma \sim g^2 \ln(g^{-1})\, T$, when extrapolated to larger coupling.
More surprising is the behavior near $T_c$: here the width has to become small since the (lattice) entropy is small. Within the present approach, this is due to the logarithm in equation (\ref{eq: gamma}) and the fit value $c \approx g^2(T_c)$, i.\,e., $\gamma_g$ vanishes almost exactly at $T_c$. This is supposedly related to the expected critical slowing down near a phase transition. It was checked in \cite{Peshi04} that the same behavior also emerges when, in the parameterization of $\gamma$, the logarithm is replaced by a factor $\exp( -\tilde c g)$, which should reflect better the physics of heavy excitations near $T_c$.

\begin{figure}[t]
 \hskip -1em
 \includegraphics[scale=0.75]{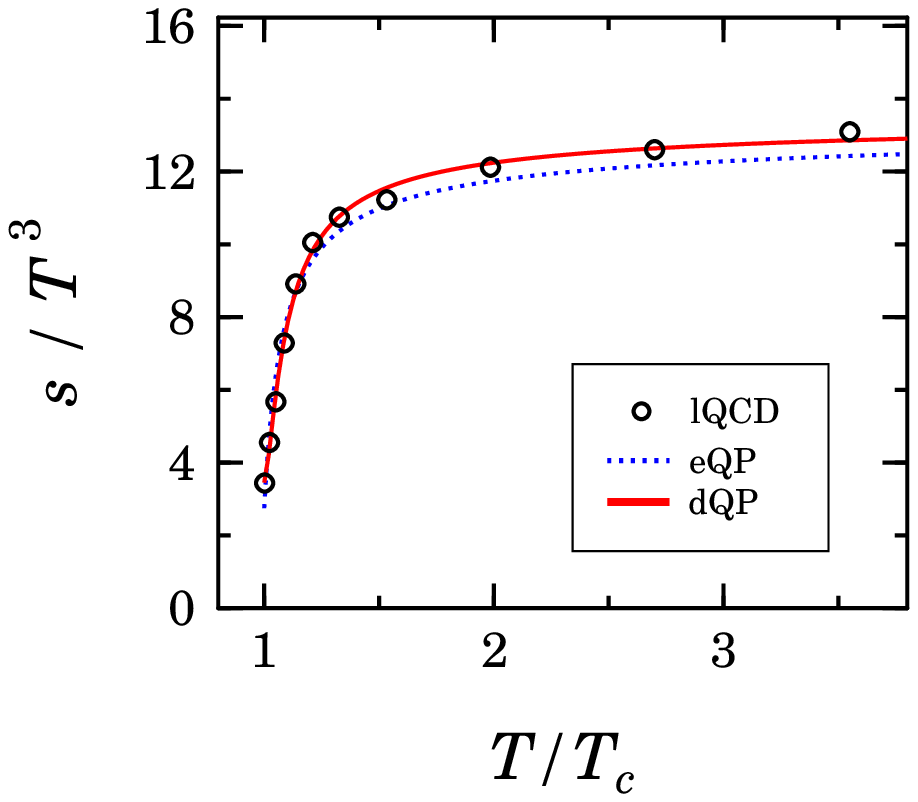}
 \hfill
 \includegraphics[scale=0.75]{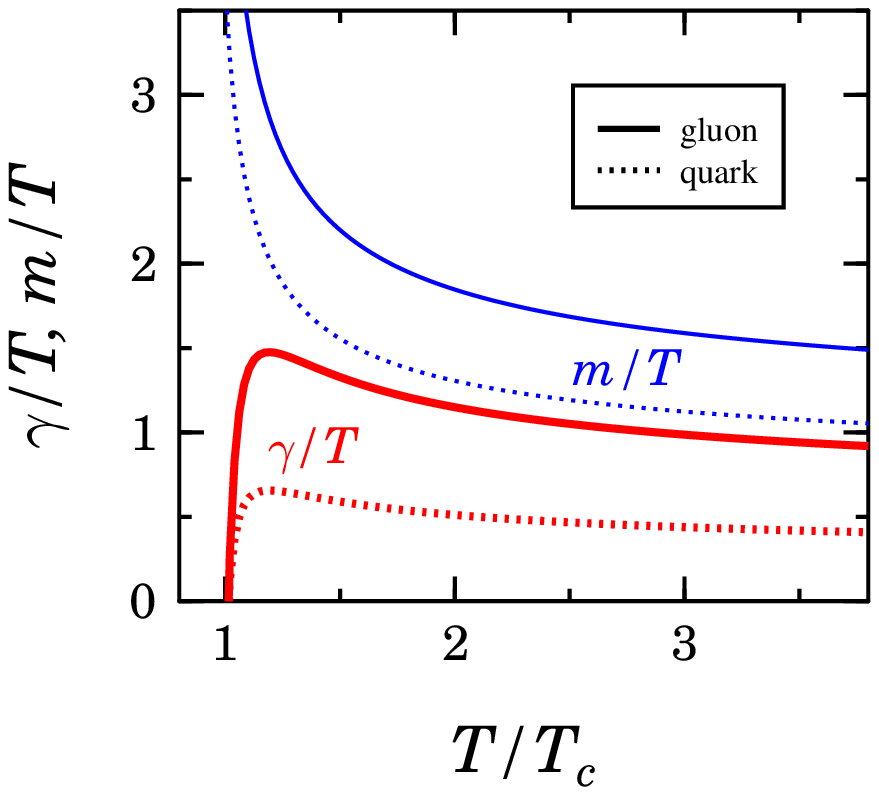}
 \caption{
    Left: the entropy $s^{QCD}$ for $\Nf = 2$ in the effective and in the 
    dynamical quasiparticle approach; the symbols represent the lattice data
    \cite{KarscLP}.
    Right: the parton masses and widths from the dynamical quasiparticle
    approach.
  \label{fig: sQCD_fitNf2}}
\end{figure}

From the rather universal behavior of the QCD entropy \cite{KarscLP} for various numbers of quark flavors (when plotted as a function of $T/T_c$ and scaled by the free limit) one can expect a similar picture for $\Nf \not= 0$ quark flavors. Figure~\ref{fig: sQCD_fitNf2} demonstrates for $\Nf = 2$ that this is indeed the case. 
I note that the lattice data are less precise than for quenched QCD; due to finite size effects the absolute scaling of the entropy data should be reduced by about 15\% \cite{KarscLP}, i.\,e., by a factor of $\tilde d \approx 0.87$. To take this into account, I allow for a rescaling of the entropy (\ref{eq: s_QCD}) by a fit factor $d$. The parameters in table \ref{tab: parms} show that for the dynamical quasiparticle approach, which gives the better fit, $d$ is close to the expected value $\tilde d$. In any case, even fixing $d = 1$ does not chance significantly the behavior of the masses and widths.
\begin{table}[b]
\centerline{
\begin{tabular}{|c|c|c|c|c|} \hline
 $\Nf = 2$               &  $\lambda$ & $T_s/T_c$ & $c$  & $d$   
\rr \\ \hline
 effective QP (no width)   &  5.1     &  0.77     &  --  & 0.97  
\rr \\ \hline
 dynamical QP (with width) &  3.7     &  0.67     & 33.6 & 0.90  
\rr \\ \hline
\end{tabular}}
\caption{The fit parameters for $\Nf = 2$, for the two quasiparticle
     approaches.
    \label{tab: parms}}
\end{table}
In principle, the parameter $c$ in equation (\ref{eq: gamma}) could be different for gluons and quarks, however, this additional freedom has again almost no effect in the fits. In particular the observation that the widths vanish at $T_c$ is robust.

\section{Implications}

\begin{figure}[ht]
 \centerline{\includegraphics[scale=0.85]{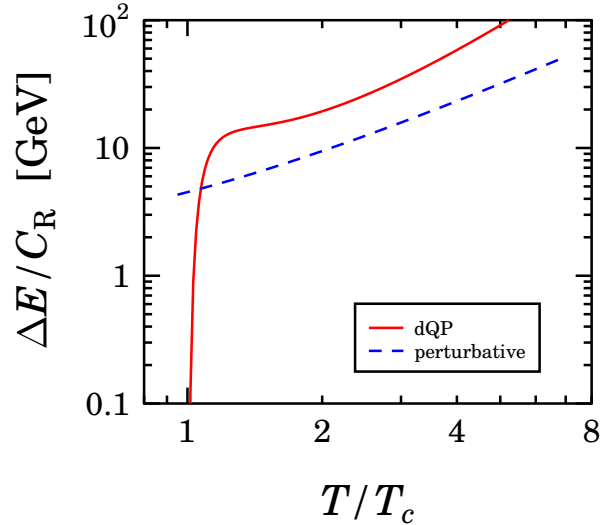}}
 \caption{The energy loss (\ref{eq: Delta E}) for $L = 5\,$fm.
    For the result denoted by `dQP', $m_D,\, \lambda_p$ and $\hat q$ were 
    taken as extracted from the analysis of quenched lattice data, 
    cf.~\cite{Peshi04}, with $T_c$ rescaled to 160\,MeV. 
    The `perturbative' estimate is based on the 2-loop running coupling,
    the perturbative result $m_D^{lo} = gT$ and a constant $\lambda_p = 
    0.3\,$fm, similar to \cite{BDMPS}.
  \label{fig: eLoss}}
\end{figure}

An interesting implication of the characteristic behavior of the width, as obtained above, is closely related to the heavy-ion experiments at SPS, RHIC and LHC. One of the mechanisms to probe the state of matter produced is the radiative energy loss of hard particles passing through the medium, in particular in the Landau-Pomeranchuk-Migdal regime of multiple coherent scattering.
Considered here for a qualitative argument is the total energy loss of a hard parton in a hot quark-gluon plasma of extension $L$ \cite{BDMPS},
\be
  -\Delta E
  =
  \textstyle\frac18\, C_R\, \alpha \hat q L^2 \ln L/\lambda_p \, ,
  \label{eq: Delta E}
\ee
where $C_R$ denotes the color representation of the jet. The properties of the medium enter mainly by the transport coefficient $\hat q = m_D^2/\lambda_p$, where the mean free path is related to the collisional width, $\lambda_p = \gamma^{-1}$, and $m_D$ is the Debye screening mass.
With the parameters $\{ \lambda, T_s/T_c, c \}$ as determined above, the temperature dependence of the coupling $\alpha = g^2/(4\pi)$ and of the mean free path are known.
The screening mass can be calculated on the lattice; for quenched QCD it was noted that the data \cite{NakaSS} can be empirically parameterized by $m_D \approx 2.7\gamma$ \cite{Peshi04}.
Thus, all the quantities determining the energy loss are adjusted in the nonperturbative regime near $T_c$, as relevant for the experiments. This leads to considerable differences to estimates based on the extrapolation of perturbative results. In particular the `critical' behavior of the width and the screening mass reflects itself in a characteristic temperature dependence of the energy loss \cite{Peshi04}, cf.~figure~\ref{fig: eLoss}.
The predicted threshold behavior could explain a quark-gluon plasma created at SPS just above $T_c$ -- without seeing partonic energy loss -- while at RHIC, reaching higher temperatures, jet quenching is effective.
\\[1mm]
To summarize, it has been proposed to infer, in the nonperturbative regime, essential 1-particle properties from quantities which can be reliably calculated, e.\,g.\ within lattice QCD. Specifically, it was shown that the small QCD entropy near $T_c$ constrains the hard parton width, while for $T > 1.2T_c$ (up to rather large temperatures) the width is of the same order as the typical energy. This behavior should have observable consequences, as argued here for the energy loss of hard jets in a hot QCD medium.

\ack{I thank the organizers of {\em Hot Quarks 04} for an inspiring conference, and BMBF for the support of this work as well as DFG for a travel support.}


\Bibliography{99}

\bibitem{Peshi04}
  A.~Peshier, Phys.\ Rev.\ D70 (2004) 034016.

\bibitem{LuttiW}
  J.\,M.\ Luttinger, J.\,C.\ Ward, Phys.\ Rev.\ 118 (1960) 1417.

\bibitem{Baym}
  G.~Baym, Phys.\ Rev.\ 127 (1962) 1391.

\bibitem{BlaizIR}
  J.\,P.~Blaizot, E.~Iancu, A.~Rebhan, Phys.\ Rev.\ D63 (2001) 065003.

\bibitem{Peshi02}
  A.~Peshier, J.~High Energy Phys.\ 01 (2003) 040.

\bibitem{BlaizI}
 J.\,P.~Blaizot, E.~Iancu, Phys.\ Rev.\ D55 (1997) 973.

\bibitem{Peshi01}
  A.~Peshier, Phys.\ Rev.\ D63 (2001) 105004.

\bibitem{eQP}
  A.~Peshier, B.~K{\"a}mpfer, O.\,P.~Pavlenko, G.~Soff,
  Phys.\ Rev.\ D54 (1996) 2399;
  A.~Peshier, B.~K{\"a}mpfer, G.~Soff, Phys.~Rev.~C61 (2000) 045203,
  Phys.~Rev.~D66 (2002) 094003.

\bibitem{Pisar93}
 R.\,D.~Pisarski, Phys.\ Rev.\ D47 (1993) 5589.

\bibitem{KarscLP}
  F.~Karsch, E.~Laermann, A.~Peikert, Phys.~Lett.~B478 (2000) 447.

\bibitem{BDMPS}
 R.~Baier, Yu.\,L.~Dokshitzer, A.\,H.~Mueller, S.~Peigne, D.~Schiff,
 Nucl.\ Phys.\ B483 (1997) 291.

\bibitem{NakaSS}
 A.~Nakamura, T.\ Saito, S.\ Sakai, Phys.\ Rev.\ D69 (2004) 014506.

\endbib
\end{document}